\documentclass[letter,11pt]{article}

\usepackage[margin=1in]{geometry}
\usepackage{color}
\usepackage{caption}
\usepackage{amsmath}
\usepackage{graphicx}
\usepackage{epstopdf}

\begin{document}

\begin{titlepage}
\title{Diffusion Component Analysis: \\ Unraveling Functional Topology in Biological Networks}
\author{Hyunghoon Cho$^1$, Bonnie Berger$^{1,2,*}$ and Jian Peng$^{1,2,3,*}$}
\date{}
\maketitle
\setcounter{page}{0}

{\small
\noindent{$^{1}$ Computer Science and Artificial Intelligence Laboratory, MIT, Cambridge, MA, USA\\}
\noindent{$^{2}$ Department of Mathematics, MIT, Cambridge, MA, USA \\}
\noindent{$^{3}$ Department of Computer Science, University of Illinois at Urbana-Champaign, Champaign, IL, USA \\}
\noindent{$^{*}$ Corresponding authors: \textit{bab@mit.edu} and \textit{jianpeng@illinois.edu}\\}
}

\thispagestyle{empty}
\abstract{
Complex biological systems have been successfully modeled by biochemical and genetic interaction networks, typically gathered from high-throughput (HTP) data. These networks can be used to infer functional relationships between genes or proteins. Using the intuition that the topological role of a gene in a network relates to its biological function, local or diffusion-based ``guilt-by-association'' and graph-theoretic methods have had success in inferring gene functions. Here we seek to improve function prediction by integrating diffusion-based methods with a novel dimensionality reduction technique to overcome the incomplete and noisy nature of network data. In this paper, we introduce diffusion component analysis (DCA), a framework that plugs in a diffusion model and learns a low-dimensional vector representation of each node to encode the topological properties of a network. As a proof of concept, we demonstrate DCA's substantial improvement over state-of-the-art diffusion-based approaches in predicting protein function from molecular interaction networks. Moreover, our DCA framework can integrate multiple networks from heterogeneous sources, consisting of genomic information, biochemical experiments and other resources, to even further improve function prediction. Yet another layer of performance gain is achieved by integrating the DCA framework with support vector machines that take our node vector representations as features. Overall, our DCA framework provides a novel representation of nodes in a network that can be used as a plug-in architecture to other machine learning algorithms to decipher topological properties of and obtain novel insights into interactomes.\footnote{This paper was selected for oral presentation at RECOMB 2015 and an abstract is published in the conference proceedings.}}
\end{titlepage}

\section{Introduction}
Cellular processes are executed through complex interactions between molecules, such as proteins, DNA, RNA and small compounds.
Recently-developed high-throughput (HTP) experimental techniques, such as yeast two-hybrid screens and genetic interaction assays, have helped to identify molecular interactions in bulk and enabled researchers to piece together large-scale interaction networks from these data 
\cite{Deane2002, Phizicky1995, Roguev2007}.

Protein-protein interaction and genetic interaction networks are the largest and most commonly-used datasets to study how molecules and their interactions determine the function of biological processes.
However, networks generated from such data are often incomplete, noisy and difficult to interpret.
Thus, we seek to design efficient algorithms to obtain accurate and comprehensive functional annotations of genes and proteins in these less-than-perfect networks.

Graph-theoretic methods have been proposed for the functional annotation of genes which use the intuition that interacting genes close together in network topology are more likely to perform similar functions 
\cite{Cao2014,Mostafavi2008,Nabieva2005,Wang2013,Wang2012, Chua2006,Milenkoviae2008,Karaoz2004}.
A type of random-walk diffusion algorithm, also known as random walk with restart (RWR), has been extensively studied in the context of biological networks and effectively applied to protein function prediction and enrichment analysis
\cite{Cao2014, Glaab2012, Kohler2008, Liao2009, Navlakha2010}.
The key idea behind such algorithms is to propagate information along the network, in order to exploit both direct and indirect linkages between genes.
Typically, a distribution of topological similarity is computed for each gene, in relation to other genes in the network, so that researchers can select the most related genes in the resulting distribution 
\cite{Glaab2012, Kohler2008, Liao2009, Navlakha2010}
or, rather, select genes that share the most similar distributions 
\cite{Cao2014}.
Such diffusion-based methods have been shown to be useful for protein function prediction, mainly due to their ability to capture fine-grain topological properties, which can be less clear in the direct interactome neighborhood of a gene. 
Nevertheless, diffusion algorithms are still far from satisfactory, partially due to the noisy and incomplete nature of high-throughput data. Even the yeast protein interactome, the highest quality data gathered for any organism, includes a large number of false positive and false negative interactions 
\cite{Deane2002, Collins2007, Gavin2006, Krogan2006}.

Dimensionality reduction techniques such as principle component analysis (PCA) have been effectively used for de-noising and improving accuracy in high dimensional biological data.
The purpose of PCA is to reduce the dimensionality of a dataset by linearly projecting it onto a lower dimensional space, while retaining most of the original dataset's variance. 
In many predictive machine learning applications, using PCA to de-noise the input data has been shown to be effective because it makes the model more resistant to overfitting.
However, to our best knowledge, little effort has been made in the spirit of dimensionality reduction and de-noising for improving diffusion-based function prediction approaches for biological networks.
Here we seek to improve function prediction by integrating diffusion-based methods with a novel dimensionality reduction technique, specifically designed to capture the diffusion patterns of a network, to overcome the incomplete and noisy nature of network data. 

In this paper, we propose Diffusion Component Analysis (DCA), a novel analytical framework that combines diffusion-based methods and dimensionality reduction to better extract topological network information in order to facilitate accurate functional annotation of genes or proteins.
The key idea behind DCA is to obtain informative, but low-dimensional features, which encode the inherent topological properties of each node in the network.
We first run a diffusion algorithm on a molecular network to obtain a distribution for each node which captures its relation to all other nodes in the network.
We then approximate each of these distributions by constructing a multinomial logistic model, parametrized by low-dimensional feature vectors, for each node.
Feature vectors of all nodes are obtained by minimizing the Kullback-Leibler (KL) divergence (relative entropy) between the diffusion and parameterized-multinomial logistic distributions.
Akin to PCA, which reveals the internal low-dimensional linear structure of the data that best explains the variance, DCA computes a low-dimensional vector-space representation for all nodes in the network such that the diffusion in the network can be best explained.
Moreover, DCA can be naturally extended to integrate multiple heterogeneous networks by performing diffusion on separate networks and jointly optimizing feature vectors.

To evaluate the performance of DCA, we first applied DCA to predict protein function in yeast molecular interaction networks and achieved substantial improvement over existing methods.  We next demonstrated the integrative capacity of DCA by using it to integrate heterogeneous networks based on genomic information, biochemical experiments and other resources from the STRING database and further improved functional annotation. In addition, we trained a support vector machine with our node-vector representation as features to be used as a metric for ``distanceÓ between genes for improved protein function prediction, as SVMs performed better than k-nearest neighbor search. Overall, when tested on the STRING networks with third level functional annotations from MIPS, our DCA framework is able to achieve 71.29\% accuracy, which is remarkably 12.31\% higher than the previous state-of-the-art, diffusion-based method. Low-dimensional feature vector representation learned by DCA can readily be plugged into other existing machine learning algorithms to decipher functional properties of and obtain novel insights into interactomes.

\section{Diffusion Component Analysis}

\begin{figure}
  \centering
  \includegraphics[width=0.9\textwidth]{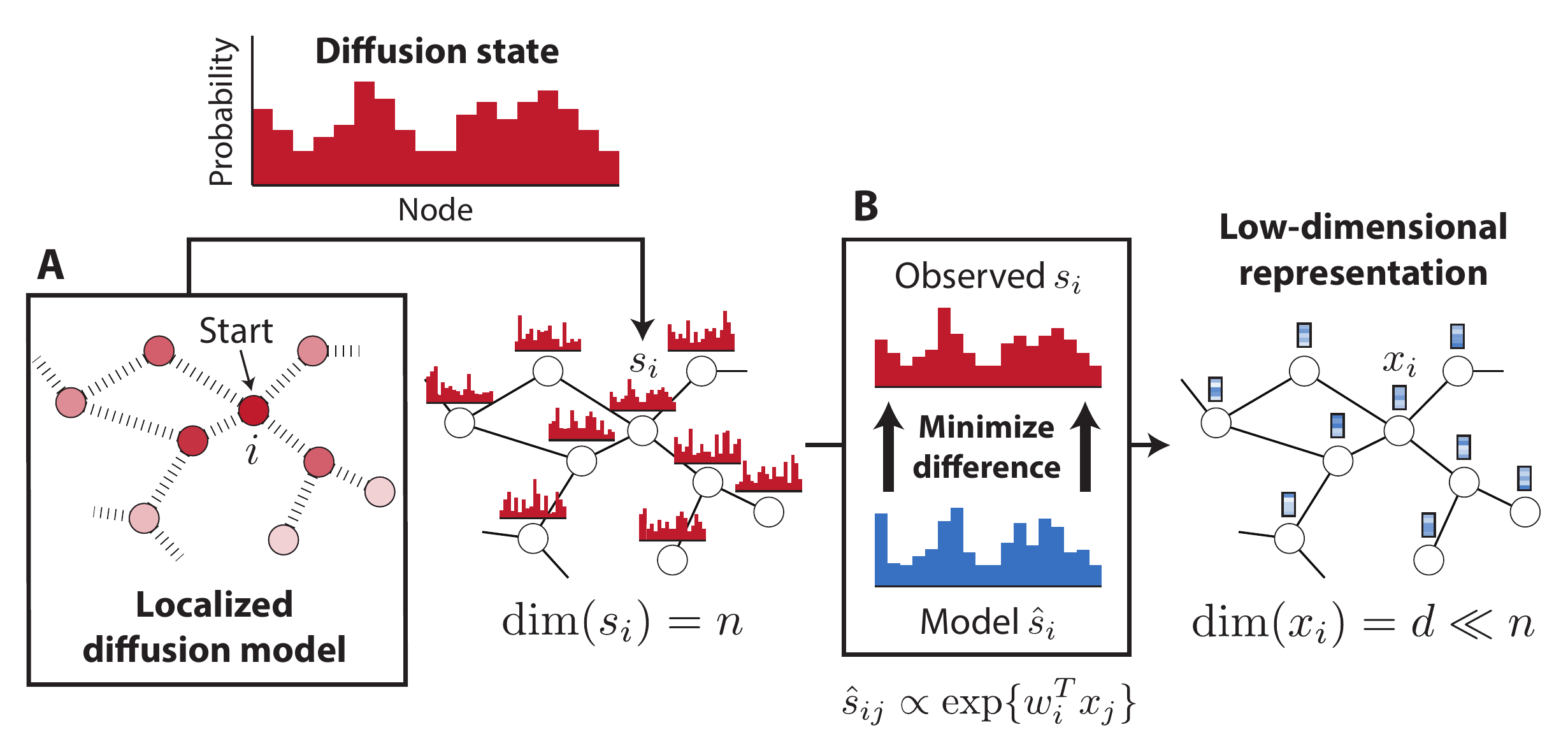}
  \captionsetup{width=0.9\textwidth}
  \caption{\small {\bf Diffusion Component Analysis.} (A) Random walks with restart are used to compute the diffusion state for each node in a network. (B) Low-dimensional feature vectors describing the topological properties of each node are obtained by minimizing the average KL-divergence between the observed diffusion states and the parameterized-multinomial logistic distributions.  }
  \label{fig:dca}
\end{figure}

Biological networks provide a useful topological structure over which information about certain nodes in a graph can be propagated to shed light on the properties of nodes that are less characterized. This relies on the assumption that the nodes that are ``similar'' or ``close'' to each other in the graph will tend to have similar properties, which is intuitively the case for biological functions or disease associations, where localized groups of interacting genes or proteins give rise to a specific molecular phenotype. Conventional approaches to finding nodes that are topologically associated with a given node involve either directly taking the neighborhood in the graph, or using localized diffusion models to select top nodes in the resulting distribution (which we refer to as \emph{diffusion state}), or to select nodes with similar distributions. The diffusion-based methods are shown to be effective in various settings, mainly due to their ability to capture fine-grain topological properties that lie beyond the direct neighborhood of a node.  Our DCA framework substantially advances previous diffusion-based methods by providing a way to extract topological information encoded in the diffusion states using a compact, low-dimensional vector representation of nodes in a graph (Figure \ref{fig:dca}).

\subsection{Random walk with restart review}
The random walk with restart (RWR) method has been well-established for analyzing network structures. By allowing the restart of random walk in each step with a probability, RWR can take into consideration both local and global topology within the network to identify the relevant or important nodes in the network. Let $A$ denote the adjacency matrix of a (weighted) molecular interaction network $G=(V,E)$ with $n$ nodes, each denoting a gene or a protein. Each entry $B_{i,j}$ in the transition probability matrix $B$, which stores the probability of a transition from node $i$ to node $j$, is computed as
\begin{equation*}
B_{i,j} = \frac{A_{i,j}}{\sum_{j'} A_{i,j'}}.
\end{equation*}
Formally, the RWR from a node $i$ is defined as:
\begin{equation}
s^{t+1}_{i} = (1-p_r)s^t_{i} B + p_r e_i
\end{equation}
where $p_r$ is the probability of restart, controlling the relative influence of local and global topological information in the diffusion, with higher chances of restart placing greater emphasis on the local structure;  $e_i$ is a $n$-dimensional distribution vector with  $e_i(i)=1$ and $e_i(j)=0, \forall j\neq i$; $s^t$ is a $n$-dimensional distribution vector in which each entry holds the probability of a node being visited after $t$ steps in the random walk. The first term in the above update corresponds to following a random edge connected to the current node, while the second term corresponds to restarting from the initial node $i$. At the fixed point of this iteration we obtain the stationary distribution $s^{\infty}_i$. Consistent with a previous work \cite{Cao2014}, we define the \emph{diffusion state} $s_i = s^{\infty}_i \in \Delta_n$ of each node $i$ to the stationary distribution of random walk with restart (RWR) starting at each node,  where $ \Delta_n $ denotes the $n$-dimensional probability simplex. Intuitively, the $j$th element, $s_{ij}$, represents the probability of RWR starting at node $i$ ending up at node $j$ in equilibrium. When the diffusion states of two nodes are close to one another, it implies that they are in similar positions within the graph with respect to other nodes, which might suggest functional similarity. This insight provided the basis for several diffusion-based methods 
\cite{Cao2014, Glaab2012, Macropol2009, Wang2012}
that aim to predict characteristics of genes or proteins by using the diffusion states to better capture topological associations. Instead of simply using the probability in the diffusion state, the diffusion state distance (DSD) approach, using $L1$ distances between diffusion states, achieved the state-of-the-art performance on predicting protein functions on yeast interactomes 
\cite{Cao2014}.

\subsection{Novel dimensionality reduction}
A key observation behind our approach is that the diffusion states obtained in this manner are still noisy, in part due to the low quality and high dimensionality of the original network data. Just a few missing or spurious interactions in the network can have significant impact on the result of random walks, and the fact that most biological networks are incomplete greatly exacerbates this problem. Moreover, the high dimensionality of the diffusion state makes it not readily useful as topological features for network-based classification or regression tasks. With the goal of noise and dimensionality reduction, we approximate each diffusion state $s_i$ with a multinomial logistic model based on a latent vector representation of nodes that uses far fewer dimensions than the original, $n$-dimensional state. Specifically, we compute the probability assigned to node $j$ in the diffusion state of node $i$ as
\begin{equation}
\hat{s}_{ij} := \frac{\exp\{w_i^T x_j\}}{\sum_{j'} \exp\{w_i^T x_{j'}\}},
\end{equation}
where $\forall i, w_i,x_i \in {\bf R}^d$ for $d \ll n$. Each node is given two vector representations, $w_i$ and $x_i$. We refer to $w_i$ as the context feature and $x_i$ as the node feature of node $i$, both capturing the intrinsic topological properties in the network. If $w_i$ and $x_j$ are close in direction and with large inner product, node $j$ should be frequently visited in the random walk starting from node $i$. Ideally, if the vector representation $w$ and $x$ is able to capture fine-grain topological properties, we can use it to retrieve genes with similar functions or use it as features for other network-based machine learning applications. While it is possible to enforce equality between these two vectors, decoupling them leads to a more manageable optimization problem and also allows our framework to be readily extended to the multiple network case, which is further described in the next section.

Given this model, we formulate the following optimization problem that takes a set of observed diffusion states $s=\{s_1,\dots,s_n\}$ as input and finds the low-dimensional vector representation of nodes $w$ and $x$ that best approximates $s$ according to the multinomial logistic model. To obtain $w$ and $x$ for all nodes, we use KL-divergence (relative entropy) as the objective function, which is a natural choice for comparing probability distributions, to guide the optimization:
\begin{equation}
\underset{w,x}{\text{minimize}}\  C(s, \hat{s})= \frac{1}{n} \sum_{i=1}^{n} D_\text{KL} (s_i \| \hat{s}_i).
\end{equation}
By writing out the definition of relative entropy and $\hat{s}$, we can express the objective as
\begin{equation}
C(s,\hat{s}) = \frac{1}{n} \sum_{i=1}^n \left[ H(s_i) - \sum_{j=1}^n s_{ij} \left( w_i^T x_j - \log \left( \sum_{j' = 1}^n \exp \{ w_i^T x_{j'} \} \right) \right)  \right],
\end{equation}
where $H(\cdot)$ denotes the entropy. Similar to PCA which finds the low-dimensional linear structure of the data to best explain the data variance, DCA computes a low-dimensional vector-space representation for all nodes in the network data such that the diffusion on the network can be best explained.

To optimize this objective function, we computed the gradients with respect to the parameters $w$ and $x$:
\begin{align}
\nabla_{w_i} C(s,\hat{s}) &= \frac{1}{n} \sum_{j=1}^n \left( \hat{s}_{ij} - s_{ij}\right) x_j, \\
\nabla_{x_i} C(s,\hat{s}) &= \frac{1}{n} \sum_{j=1}^n \left( \hat{s}_{ji} - s_{ji}\right) w_j.
\end{align}
Both the objective function and the gradients can be computed in $O(n^2d)$ runtime complexity. We use a standard quasi-Newton method L-BFGS 
\cite{Zhu1997}
with these gradients to find the low-dimensional vector representations $w$ and $x$ corresponding to a local optimum of this optimization problem. We use uniform random numbers from $[-0.05,0.05]$ to initialize the vectors and observed that the optimization can almost surely find good local optimal solutions.

\subsection{Novel integration of heterogeneous networks}
\label{sec:integration}
Our framework gives rise to a novel approach to integrating heterogeneous data sources. As huge amount of interaction data generated from wide variety of experimental and computational studies for molecular networks, integration approaches have been used to combine them to a network model and identify genes' functions 
\cite{Mostafavi2008,Wang2012,Pena2008,Hwang2011,Lee2004,Wang2014}.
A common approach taken by previous work is based on weighting the networks in some manner to get an integrated network before analyzing the topological properties of each node \cite{Mostafavi2008,Lee2004,Franceschini2013}.
For example, in the STRING database, confidence scores from heterogeneous sources including co-expression relationship, genomic context, experimental interactions and literature evidence, are combined by the Bayes method: $p_{i,j} = 1 - \prod_k (1-p^{(k)}_{i,j})$, where $p^{(k)}_{i,j} \in [0,1]$ is the probability confidence of interaction $(i,j)$ from network $k \in \{1,2,...,K\}$ derived by the same data source. However, by collapsing the networks into a single representative network, we lose important network-specific patterns. Performing random walk directly on the integrated network can be problematic, since edges from multiple sources are mixed.

Here, we naturally extend our DCA approach to integrate network data from diverse sources. We first perform random walks on each individual network $k$ separately and obtain network-specific diffusion states $s^{(k)}_i$ for each node $i$. We also construct the multinomial distribution $\hat{s}^{(k)}_{ij}$ from the following logistic model,
\begin{equation}
\hat{s}^{(k)}_{ij} := \frac{\exp\{w^{(k)T}_i x_j\}}{\sum_{j'} \exp\{w^{(k)T}_i x_{j'}\}},
\end{equation}
where for each node $i$ in network $k$, we assign it a network-specific context vector representation $w^{(k)}_i$, which also encodes the intrinsic topological properties of network dataset $k$; for node features $x$, we allow them to be shared across all $K$ networks. Finally, we jointly optimize the following objective function,  
\begin{equation}
\underset{w,x}{\text{minimize}}\  C(s, \hat{s})= \sum_{k=1}^{K} \frac{1}{n} \sum_{i=1}^{n} D_\text{KL} (s^{(k)}_i \| \hat{s}^{(k)}_i),
\end{equation}
by the quasi-Newton L-BFGS method. Note that it is possible to weight the divergence term of each network differently, but we give equal importance to each network in this work for simplicity.

\subsection{Evaluation of DCA in protein function prediction}
To assess the quality of vector representations obtained by DCA, we considered the task of protein function prediction from protein interaction networks. In protein-protein interaction networks, functional relationship between proteins is believed to correlate with the  proximity among proteins in the network structure, since interacting proteins are often involved within the same biological processes. Thus using functional annotations from well characterized proteins to predict the function of unlabeled proteins can be effective if we have a good way to compute the topological proximity or similarity between proteins. Given a distance metric that well captures the topological similarity of proteins, we are able to predict function of a protein from the known functions of closest proteins ranked by this distance metric. This makes function prediction a good example to evaluate how good the topological and diffusion information are encoded by DCA.

To predict protein function using DCA, we compute the similarity between any two proteins in the network based on their low-dimensional vector representations. For a pair of proteins $i$ and $j$, we used the cosine distance between their node features as following.
\begin{equation}
D_{cos} (i,j) = 1-\frac{x_i^T x_j}{\|x_i\| \|x_j\|}
\end{equation}
After the distances are computed, we use weighted majority vote with ten closest proteins to assign functions to unlabeled proteins. Unless otherwise noted, we use majority vote with cosine distances to carry out function prediction based on our vector representation of proteins.

Furthermore, since DCA gives low-rank features for each protein in the network, they can be readily used as feature vectors for machine learning algorithms in various applications, such as classification or clustering. To demonstrate the potential of DCA as a plug-in framework for existing machine learning algorithms, we formulate the protein function prediction task as a multi-label classification problem and applied an off-the-shelf support vector machine (SVM). Since there are proteins with multiple functional annotations, we train a SVM for each function category and then assign functions to a protein with annotations with highest output values from SVM. We denote this method as DCA-SVM.

\section{Results}

Here we evaluate the ability of our DCA framework to uncover functional relationships in the interactome of yeast, which is one of the few organisms for which a comprehensive annotation of proteins is available. With the combination of noise reduction via dimensionality reduction, improved integration of multiple heterogenous networks, and the use of support vector machines, our DCA framework is able to achieve 71.29\% accuracy with five-fold cross-validation on the STRING networks with third level functional annotations from MIPS, which is remarkably 12.31\% higher than the previous state-of-the-art diffusion state distance (DSD) method. The details of our experiments follow.

\subsection{Networks and annotations}

We obtained a collection of protein-protein interaction (PPI) networks of yeast from the STRING \cite{Franceschini2013} database v9.1, which is based on a variety of data sources, including high-throughput interaction assays, curated PPI databases, and conserved coexpression. We excluded the network constructed from text mining of academic literature to prevent confounding caused by links based on functional similarity. The resulting collection consisted of six heterogenous networks over 6,400 proteins, with the number of edges varying from 1,361 to 314,013. Every edge in these networks is associated with a weight between 0 and 1 representing the probability of edge presence, which we factor into the calculation of transition probabilities in the random walk process. For methods that take a single network as input, we used STRING's approach to integrate the networks: we assign $p_{i,j} = 1 - \prod_k (1-p^{(k)}_{i,j})$ as the probability of each edge in the combined network, where $p^{(k)}_{i,j}$ is the probability associated with the same edge in network $k$. 536,207 edges were produced after the integration.

Furthermore, we built a PPI network of 6,071 yeast proteins based on 135,374 physical interactions reported in the BioGRID \cite{Stark2006} database v3.2.117. We considered every edge in the database to be undirected for simplicity, although our method can be easily applied to directed networks as well. For pairs of nodes with multiple edges, we merged the redundant edges and assigned a weight corresponding to the number of merged edges to produce a weighted graph. We used this network as an additional test case for our framework.

To train and test function prediction methods, we obtained functional annotations for yeast proteins from the Munich Information Center For Protein Sequences (MIPS) Functional Catalogue \cite{Ruepp2004}. The biological functions in this database are organized in a three-layered hierarchy, where the top level (MIPS1) consists of 17 functions, the second (MIPS2) consists of 74 functions, and the third (MIPS3) consists of 154 functions. Importantly, each protein can have more than one associated function. All of our experiments were repeated for each level of annotations to test the methods in different challenging settings, with MIPS3 being the most difficult due to the largest number of candidate labels.

\subsection{Performance metrics}\label{metrics}

For each function prediction method, we repeatedly held out 20\% of the annotated proteins as the validation set and used the remaining 80\% to predict their functions. We used two prediction performance metrics that are commonly used in this field. First, we calculate the \emph{accuracy} by assigning top function to each protein in the validation set and measuring how often our prediction is one of the known functions of the protein. Second, we consider the micro-average \emph{F1 score}. To calculate this metric, we assign top $\alpha$ predictions to each protein, construct a 2-by-2 contingency table for each function by treating it as a binary classification problem, and compute the F1 score using the combined table where each entry is summed across all functions. We used $\alpha = 3$ for the results presented in this paper, following previous work \cite{Cao2014, Schwikowski2000}.

\subsection{DCA substantially improves functional annotation}

\begin{figure}[t!]
  \centering
  \includegraphics[width=1\textwidth]{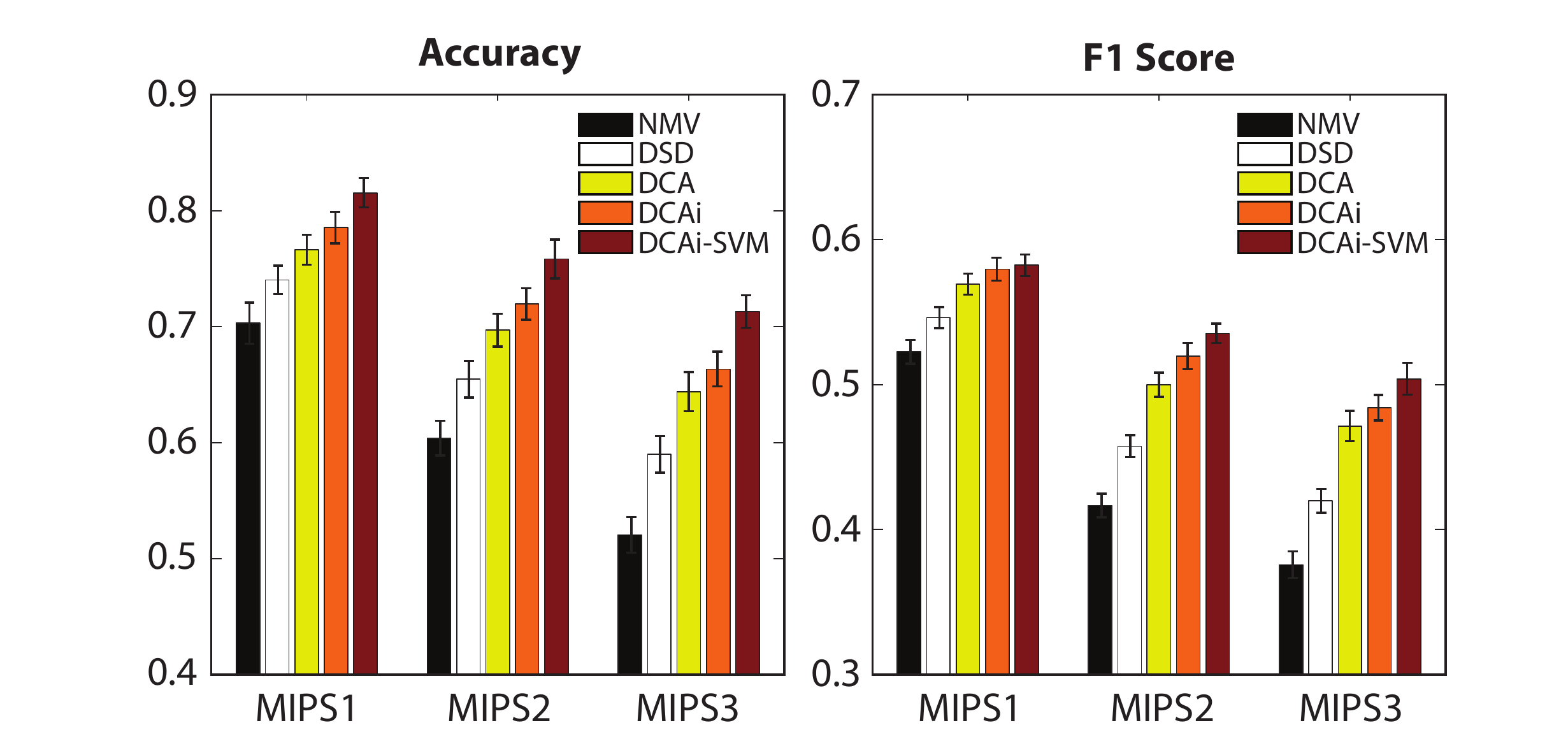}
  \caption{{\bf Protein function prediction performance on yeast STRING networks.} DCA (diffusion component analysis) has substantially greater accuracy than both NMV (neighbor majority vote) and DSD (diffusion state distance) across all three MIPS levels. We used 500 dimensions for DCA, DCAi, and DCAi-SVM.}
  \label{fig:main-comparison}
\end{figure}

The results for protein function prediction using the STRING network (without text mining) are summarized in Figure \ref{fig:main-comparison}. We used restart probability of 0.5 for random walks and obtained 500-dimensional vector representation for each node via DCA. We observed that our performance is stable for different values of restart probability between 0.5 and 0.9. We compared our framework to two other baseline methods. The first baseline is \emph{neighborhood majority vote} (NMV) \cite{Schwikowski2000}, in which predictions are produced by prioritizing the functions based on the number of times they tag proteins that are directly connected to the target protein. This method is not able to make use of any information associated with proteins that are more than one hop away from the target. This issue is addressed by the second baseline we considered: \emph{diffusion state distance} (DSD), which is the current state-of-the-art, diffusion-based method \cite{Cao2014}. Here, the diffusion state of each node is obtained in a similar fashion as DCA. Then for each target protein, $k$ most similar proteins, based on the L1 distance metric imposed over the diffusion states, are taken for a majority vote. In particular, each protein casts a vote for each of its associated functions with weight equal to the reciprocal of the distance to the target protein, and finally the functions are prioritized by the sum of these votes. We used $k = 10$, following the original work that introduced this method \cite{Cao2014}. The diffusion states can capture longer-range topological properties in addition to the local neighborhood, which allows DSD to achieve a large improvement in prediction performance over NMV, as can be seen in Figure \ref{fig:main-comparison}. However, through our DCA frameworkÕs substantial advances, it achieves a significant improvement over DSD on all three levels of MIPS (Figure \ref{fig:main-comparison}). This result highlights that, by reducing each diffusion state into a low-dimensional vector using DCA, one can capture functional associations between proteins more accurately.

One of the strengths of DCA is its ability to easily and principally integrate topological information from multiple networks. This integration is achieved by learning a canonical vector representation of proteins that explains their diffusion patterns across all networks, while simultaneously capturing network-specific effects using context-specific feature vectors (see \ref{sec:integration}). Intuitively, this approach is able to capture more fine-grain topological patterns which would not be visible if we were to flatten the networks into a single representative network, as most do, before analyzing the topology. By integrating the STRING networks with our approach and using the canonical vectors to guide the majority vote for function prediction based on pairwise cosine distances (denoted as DCAi), we achieve improvement on all levels of MIPS on top of DCA with STRING's integration (Figure \ref{fig:main-comparison}). This result suggests that functional relations can be better explained by individually modeling diffusion patterns within each network. \\

To demonstrate the potential of DCA as a plug-in framework for machine learning algorithms, we cast the protein function prediction task as a multi-label classification problem and applied an off-the-shelf support vector machine (SVM) toolbox, LIBSVM \cite{Chang2011}, with the node representation from DCA as input features. We used the radial basis function kernel and performed a nested five-fold cross-validation within the training data to select the optimal parameters. The candidate values for parameter selection were $\{0.5,0.25,0.125\}$ for the variance parameter $g$ and $\{0.5, 1, 2\}$ for the cost parameter $C$. With this pipeline, we trained a binary classifier for each function and obtained a per-class probability score for each protein in the validation set. Note that an SVM does not provide probability outputs by itself, but LIBSVM supports an additional routine to train a logistic model on top of the SVM output via cross-validation to estimate the probabilities. These probability scores are then used to prioritize candidate functions for each protein, after which the prediction performance was evaluated. Simply by using an off-the-shelf SVM in this manner with the vectors from DCAi (denoted as DCAi-SVM), we were able to obtain yet another substantial performance gain in function prediction on all three levels of MIPS, as shown in Figure \ref{fig:main-comparison}. While one could also apply an SVM with the original, high-dimensional diffusion states as feature representation, we found that it performs significantly worse than training the SVM on the DCA vectors with STRING's integration (for fair comparison), denoted as DCA-SVM, which suggests that DCA is able to find a higher quality feature representation of the input data using much fewer dimensions (Supplementary Table 1). Furthermore, DCA-SVM performs worse than DCAi-SVM with our novel integration, implying that each of the improvements we made is necessary to achieve the best performance overall (Supplementary Table 1).

\begin{figure}[t!]
  \centering
  \includegraphics[width=1\textwidth]{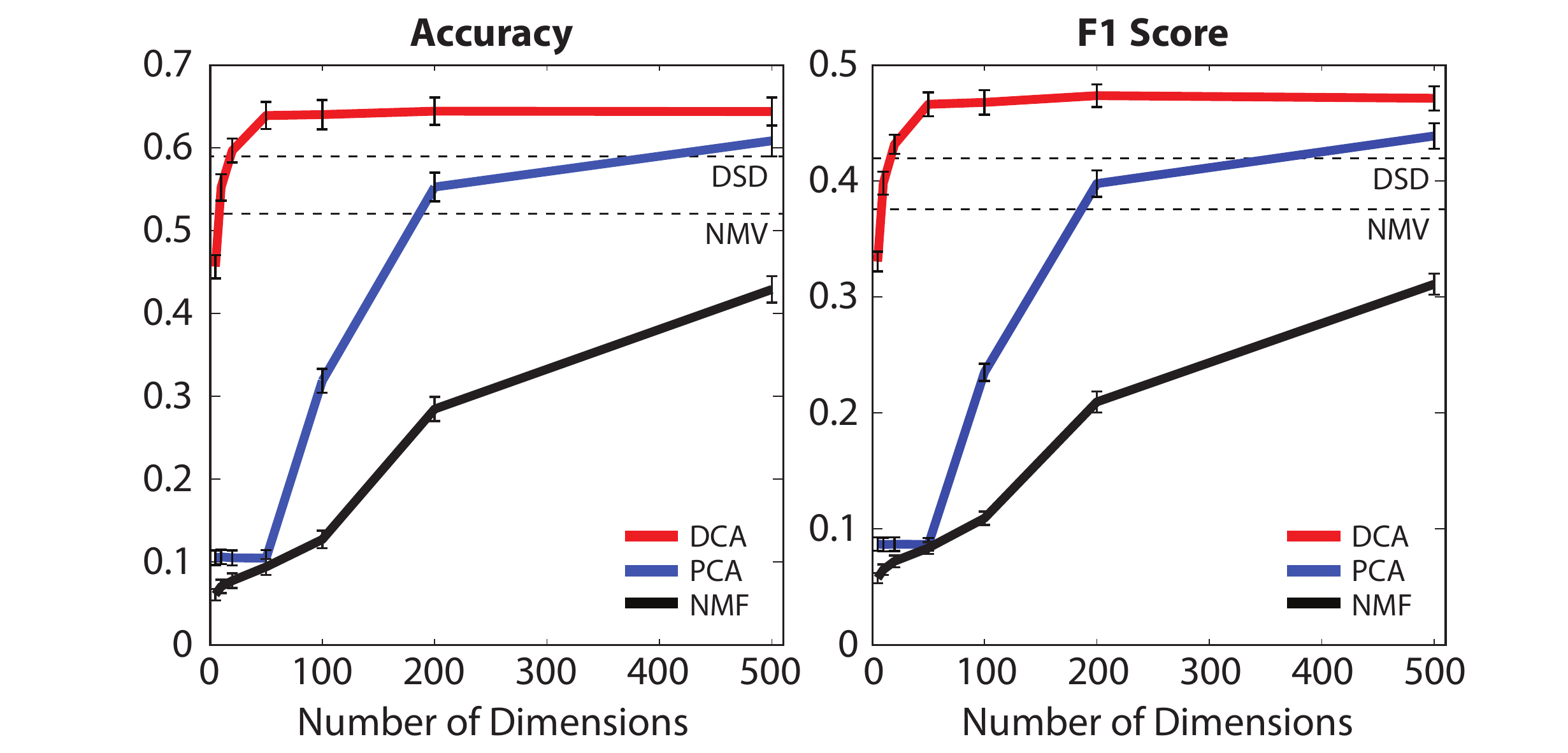}
  \caption{{\bf Comparison with other dimensionality reduction methods.} DCA (diffusion component analysis) has much greater accuracy than principal component analysis (PCA) and non-negative matrix factorization (NMF) for different numbers of dimensions, especially for fewer dimensions. The results are based on majority vote prediction where the neighborhood is defined with cosine distance. STRING networks and MIPS3 annotations are used. Baseline performances of NMV and DSD are shown as constants.}
  \label{fig:compare-dim-reduction}
\end{figure}

After observing the potential in using fewer dimensions to encode topological information, we then asked whether other conventional approaches to dimensionality reduction can achieve similar performance improvements. To this end, we compared DCA to principal component analysis (PCA) and non-negative matrix factorization (NMF) for different numbers of dimensions, and the results based on majority vote prediction based on cosine distances are summarized in Figure \ref{fig:compare-dim-reduction}. We found that DCA significantly outperforms both PCA and NMF overall. The performance gap is even more noticeable for fewer number of dimensions ($\leq 200$), which suggests that DCA is effective at extracting meaningful topological associations even in a highly constrained setting, where the conventional approaches tend to suffer, presumably due to the fact that they aim to find a linear embedding to explain variance within the data without directly modeling the diffusion states as probability distributions. In addition, our results show that the performance of DCA is stable over a wide range of values for the number of dimensions, implying that our framework is robust to overfitting. Surprisingly, DCA can achieve performance comparable to DSD, which used $n=6400$ dimensions, using only 20 dimensions. Furthermore, we found that training SVMs with feature representations from PCA and NMF perform relatively poorly, which provides additional evidence that DCA finds a good feature representation using a small number of dimensions (Supplementary Table 1).

In summary, compared to DSD, which is the current state-of-the-art function prediction method based on diffusion, we were able to achieve a 12.31\% increase in accuracy and 8.43\% increase in the micro-average F1 score (see \ref{metrics}) on MIPS3, resulting in 71.29\% accuracy and 50.40\% F1 score, by using the full DCA framework, which combines our dimensionality reduction, a novel network integration approach and SVM classifiers. This is a notable improvement, given that DSD improves upon NMV --- a naive method solely based on local topology --- by only 6.94\% in accuracy and 4.41\% in F1 score on the same data. We also observed improved performance over DSD in a different yeast PPI network, constructed only from physical interactions in the BioGRID database (Supplementary Table 2).

\section{Discussion}
We have presented Diffusion Component Analysis (DCA), a novel analytical method for biological network analysis. DCA exploits the topology of the network by diffusion and then computes a low-dimensional but highly informative vector representation for nodes in the network to approximate the diffusion information. DCA can be naturally extended to integrating multiple heterogeneous networks by first performing diffusion on separate networks and then jointly optimizing feature vectors.

We have demonstrated DCA in exploiting functional topology in molecular networks by accurately predicting protein function from molecular interaction network data. We have demonstrated substantial improvements over the state-of-the-art diffusion-based approach and other dimensionality reduction approaches.  DCA accurately encodes both local and global topological properties for all genes in the network, and thus is very useful for predicting gene functions that cannot be simply inferred from the direct interactome. Furthermore, low-dimensional vectors computed by DCA are a highly informative feature for describing the role of proteins in terms of functional topology, making them readily incorporable into existing machine learning methods.

In the future, we plan to pursue further improvements in function prediction with DCA. For example, in this work, we simply used the straightforward implementation of the supervised multilabel SVM, which is not ideal and prone to overfitting for species (e.g. Human and Mouse) with numerous functional labels but only sparsely annotated genes. We believe that functional label similarity \cite{Wang2013, Sefer2011} and gene ontology hierarchy \cite{Clark2013, Mostafavi2009, Jiang2008} can be applied with DCA to address the challenges of supervised multi-label function prediction. Other features, such as sequence or evolutionary information, which provide extra information, can be naturally combined with DCA. We also hope to explore other network-based applications, including identifying functional modules, network-guided enrichment analysis, discovering new gene ontology terms from molecular network data and network-based approaches for cancer genomics 
\cite{Glaab2012,Cornish2014,Dutkowski2013,Hofree2013,Ulitsky2007,Yosef2009}.

\bibliographystyle{unsrt}
\bibliography{dca}

\begin{figure}[p]
  \includegraphics[trim=40 0 0 20,clip,width=.85\paperwidth]{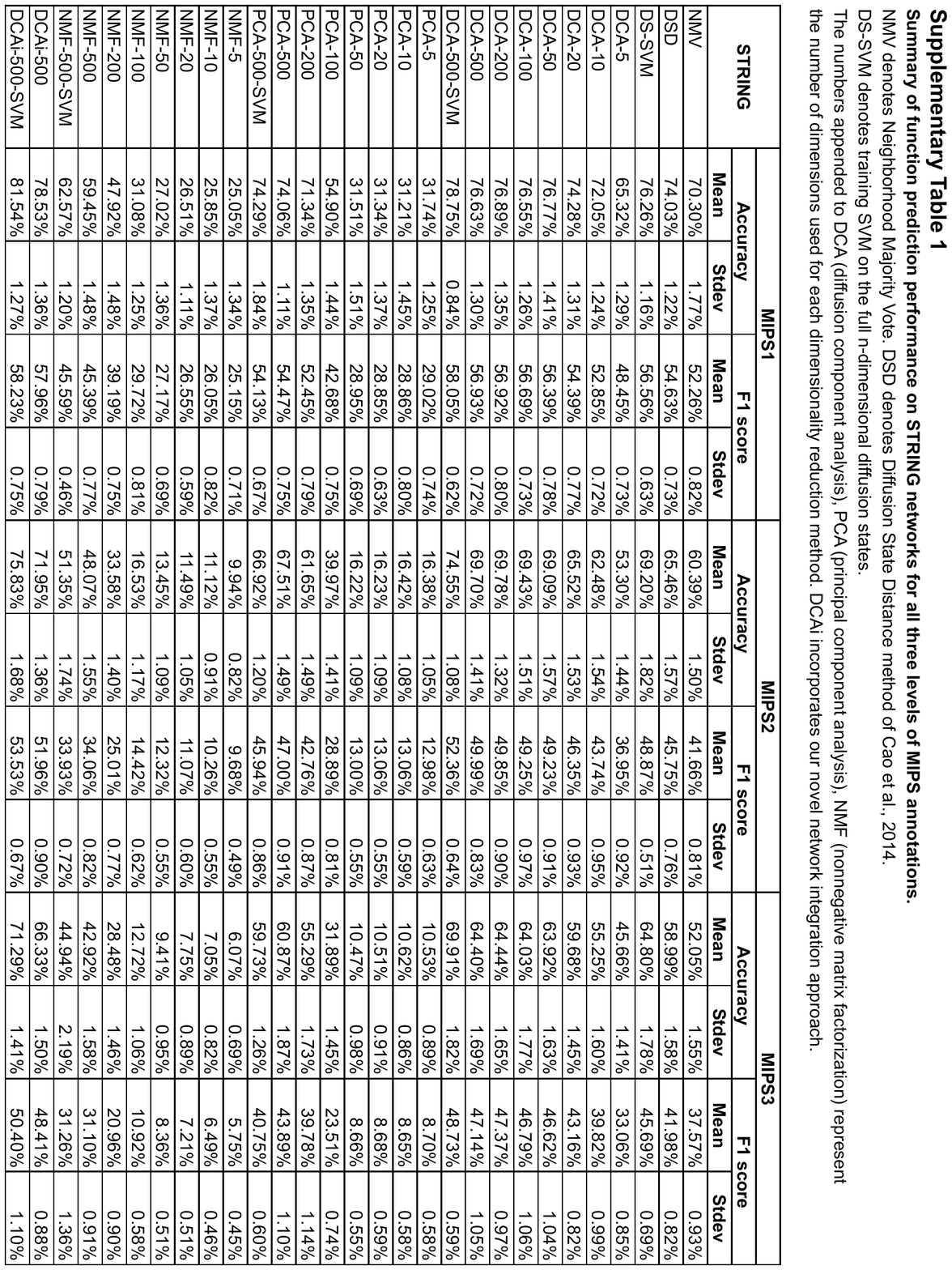}
\end{figure}

\begin{figure}[p]
  \includegraphics[trim=20 0 0 20,clip,width=.83\paperwidth]{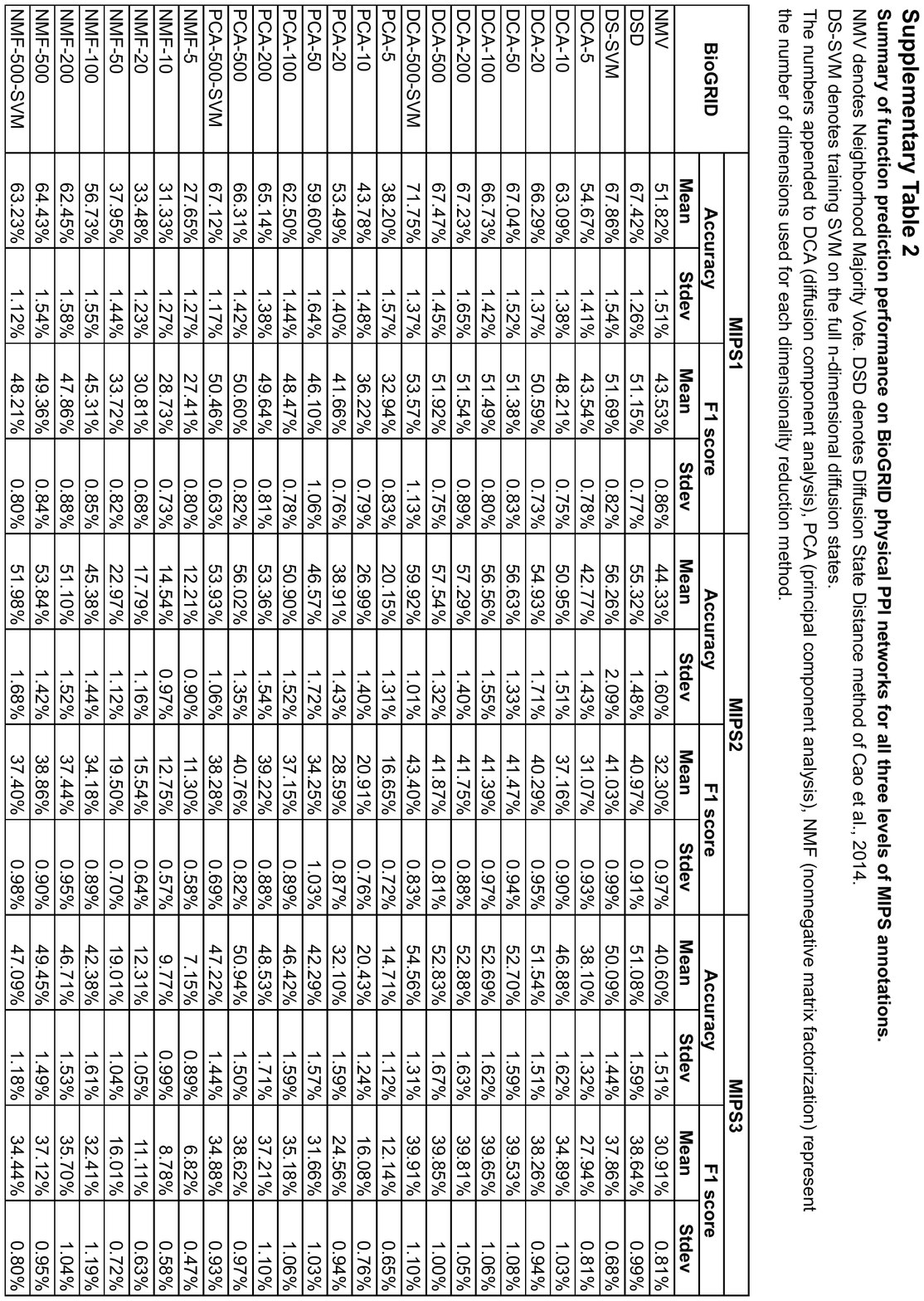}
\end{figure}

\end{document}